\documentclass[12pt]{article}
\usepackage[english]{babel}
\usepackage{amsmath}
\usepackage{color}
\Roman{section}
\newcommand{\eq}[1]{\begin{equation}#1\end{equation}}
\newcommand{\naw}[1]{\left(#1\right)}
\newcommand{\ket}[1]{\left|#1\right>}
\newcommand{\bra}[1]{\left<#1\right|}
\newcommand{\av}[1]{\left<#1\right>}

\newcommand{\modu}[1]{\left|#1\right|}
\newcommand{\poisson}[1]{\left\{#1\right\}}

\begin{document}

\begin{center}
\textsc{\Large{Quantum two-players games, entanglement and Nash equilibria}}

\emph{Katarzyna Bolonek-Laso\'n\footnote{kbolonek1@wp.pl}\\ Faculty of Economics and Sociology, Department of Statistical Methods\\ University of Lodz, Poland.}

\end{center}

\begin{abstract}
The two-players N strategies games quantized according to the Eisert-Lewenstein-Wilkens scheme \cite{EisertWL} are considered. It is shown that in the case of maximal entanglement no nontrivial pure Nash equilibrium exists. The proof relies on simple geometric properties of "chiral" group $SU\naw{N}\times SU\naw{N}$ and is based on considering the stability subgroup of the initial state of the game. The explicit forms of neither the gate operator nor the payoff matrix are necessary.  
\end{abstract}

\section{Introduction}

Since the appearance of the seminal paper of Eisert, Wilkens and Lewenstein \cite{EisertWL}, \cite{EisertW} the theory of quantum games has been a subject of intensive research \cite{Meyer}$\div$\cite{Nawaz6}. Particular attention has been paid to the problem of non-classical Nash equilibria \cite{Nash}. Once the set of admissible strategies is selected, their existence and position depend on the degree of entanglement of initial state of the game as well as the form of payoff matrix. 

The simplest and natural assumption concerning the set of acceptable strategies is to admit all unitary ones. Our aim here is to show that, once such an assumption is adopted, there exists no pure Nash equilibrium (except a trivial one) if the gate operator yields the maximally entangled initial state. The proof is based on simple group-theoretical considerations; neither the explicit form of gate operator nor that of payoff matrix are necessary.

\section{ELW games}
Let us remaind shortly the main elements of ELW quantization of classical games \cite{EisertWL}, \cite{EisertW}. We start with classical two-players two strategies game defined by the payoff matrix

\begin{center}\begin{tabular}{|c|c|c|}\hline
 \cline{1-3}
A $\setminus$ B & C & D\\ \cline{1-3}
C & (r,r) & (s,t)\\ \cline{1-3}
D & (t,s) & (p,p)\\ 
 \hline
\end{tabular}\\
\end{center}

On the quantum level one introduces two two-dimensional Hilbert spaces $H$, one for each player; the total space of the states of the game is $H\otimes H$. The basic vectors in each Hilbert space are denoted by 
\eq{\ket{C}=\left( \begin{array}{c}
1\\
0 
\end{array}\right ), \qquad \ket{D}=\left ( \begin{array}{c}
0 \\ 1 \end{array}\right ).}

The strategies of Alice and Bob are represented by unitary matrices $U_A$ and $U_B$. The final state of the game reads
\eq{\ket{\Psi_f}=\naw{J^+\naw{U_A\otimes U_B}J}\ket{C}\otimes\ket{C}}
where $J$ is a reversible two-bit gate introducing quantum entanglement.
Given $\ket{\Psi_f}$ one can compute the expected payoffs of both players. For example, the Alice payoff reads
\eq{\$_A=rP_{CC}+pP_{DD}+tP_{DC}+SP_{CD}}
with $P_{\sigma\sigma'}\equiv\modu{\av{\sigma\sigma'|\Psi_{f}}}^2$, $\sigma,\sigma'=C,D$.\\
The key point is the choice of gate operator $J$. One demands the classical strategies to be included into the quantum game; this yields
\eq{J=\exp\naw{-\frac{i\gamma}{2}\sigma_2\otimes\sigma_2}}
with $\gamma\in\av{0,\frac{\pi}{2}}$.

The properties of the game depend on:\\
(i) the choice of payoff matrix\\
(ii) the value of the $\gamma$ parameter\\
(iii) the choice of the subset $S\subset SU(2)$ of admissible strategies $U_{A,B}$.

In particular, if classical payoffs obey $t>r>p>s$, the Prisoner Dilemma emerges on the classical level. Eisert et al. \cite{EisertWL} have shown that for $\gamma=\frac{\pi}{2}$, with the proper choice of $S$, new, genuinely quantum Nash equilibrium appears (which is also Pareto optimal) allowing the players to escape the dilemma.
 
The existence of quantum Nash equilibrium depends strongly on the choice of the subset $S$. In fact, it can be shown \cite{BenjaminHay}, \cite{BenjaminHay1}, \cite{Landsburg1}, \cite{Bolonek} that for $S=SU(2)$ (i.e. all unitary strategies are admitted) and maximal entanglement, $\gamma=\frac{\pi}{2}$, to any strategy of one player there exists an appropriate counter-strategy of the second one; as a result no nontrivial Nash equilibrium exists.\\
As mentioned above, the value $\gamma=\frac{\pi}{2}$ corresponds to the maximal entanglement of the initial state 
\eq{\ket{\Psi_i}=J\naw{\ket{C}\otimes\ket{C}}.\label{a}}
Then the ELW game with all unitary strategies allowed has a number of specific properties. Apart from the existence of cunterstrategy to any given strategy it admits quaternionic description \cite{Landsburg1}, \cite{Landsburg2}, \cite{Bolonek} as well as the the one in terms of real Hilbert space \cite {Bolonek}. All these properties are related to the fact that the stability subgroup of the initial state is $SU(2)$ \cite{Bolonek}.

Let us now consider the classical game between two players, each having $N$ startegies at his/her disposal. The payoff matrix for each player is now $N\times N$ matrix. The game can be quantized following the ELW method. To each player we ascribe the $N$-dimensional Hilbert space $H$. Let $\poisson{\ket{e_i}}^N_{i=1}$ be an orthonormal basis in $H$. We put $\av{e_i|C}=\delta_{i1}$ and adopt eqs. (\ref{a}) as a definition of initial state for general $N$-strategies games; $J$ is appropriately chosen gate operator. The final state which, together with the payoff matrix, allows to compute the expected payoffs of players reads, in analogy with eq.(2),
\eq{\ket{\Psi_f}=J^+\naw{U_A\otimes U_B}J\naw{\ket{C}\otimes\ket{C}}}
where now $U_A,U_B\in SU(N)$ are the strategies of Alice and Bob; we make again the natural assumption that $S=SU(N)$ is the set of allowed strategies.

 The main point is again the choice of the gate operator $J$ \cite{Bolonek1}. As in the $N=2$ case we assume that the classical strategies are included into quantum scheme. Moreover, to leave as much freedom as possible for the choice of $J$ it is further assumed that the matrices representing classical strategies commute. As a result, $J$ depends on some, in general arbitrary, parameters $\gamma_i$,
 \eq{J=J\naw{\underline{\gamma}};}
 actually, the number of $\gamma$ parameters is $N \choose 2$ and, allowing some unitary rotation of the vector $\ket{C}$, one can represent $J$ as an exponent of linear combination of symmetrized tensor products of the Cartan subalgebra elements \cite{Bolonek1}. The explicit form of $J\naw{\gamma}$ will be not needed in what follows.
 
 \section{Nonexistence of Nash equilibria in the maximally entangled case}
 Let us assume that the parameters $\gamma$ are adjusted in such a way that the initial state (\ref{a}) is maximally entangled. We put 
 \eq{\ket{\Psi_i}=J\naw{\underline{\gamma}}\naw{\ket{C}\otimes\ket{C}}\equiv F_{ij}\ket{e_i}\otimes\ket{e_j}\label{c}}
 where the summation over repeated indices is understood. The matrix $F$ is symmetric as we are considering symmetric game. The corresponding density matrix reads 
 \eq{\rho_i= \ket{\Psi_i}\bra{\Psi_i}.}
 
 The state described by $\rho_i$ is maximally entangled if the reduced density matrices are proportional to the unit matrix \cite{Plenio}
 \eq{Tr_A\rho_i=\frac{1}{N}I,\qquad Tr_B\rho_i=\frac{1}{N}I.\label{b}}
Eqs. (\ref{b}) imply 
\eq{FF^+=\frac{1}{N}I}
i.e. the matrix $\tilde{F}\equiv\sqrt{N}F$ is unitary. \\
Let us apply the unitary transformation $U_A\otimes U_B$ to $\ket{\Psi_i}$;
\eq{\naw{U_A\otimes U_B}\ket{\Psi_i}=\naw{U_AFU^T_B}_{ij}\ket{e_i}\otimes\ket{e_j}.}
By virtue of (\ref{c}) the invariance of the initial state implies 
\eq{U_A\tilde{F}U^T_B=\tilde{F}.\label{d}}
The general solution to eq. (\ref{d}) reads
\eq{\begin{split}
& U_A=U\\
&U_B=\tilde{F}\overline{U}\tilde{F}^+.
\end{split}}
where $U\in SU(N)$ is arbitrary.
    
We conclude that the stability subgroup of $\ket{\Psi_i}$ is, up to an automorphism, the diagonal subgroup of $SU(N)\times SU(N)$.
The Lie algebra of this subgroup induces the symmetric Cartan decomposition of $ su(N)\bigoplus su(N)$.\newline The coset manifold $SU\naw{N}\times SU\naw{N}/diag\naw{SU\naw{N}\times SU\naw{N}}$ is isomorphic, as a manifold (but not as a group!), to the $SU(N)$ manifold. This allows us to write out a useful decomposition of any element of $SU(N)\times SU(N)$. Explicitly, let $U_1,U_2,V\in SU(N)$ be arbitrary; then (cf. Ref. \cite{Bolonek})
\eq{\naw{U_1,U_2}=\naw{V,U_2\tilde{F}\overline{U}^+_1\overline{V}\tilde{F^+}}\naw{V^+U_1,\tilde{F}\overline{V}^+\overline{U}_1\tilde{F}^+}.\label{e}}

The above equation has the following interpretation. Assume Alice chose an arbitrary strategy $V\in SU(N)$. Let $\naw{U_1,U_2}$ be a pair of strategies leading to the expected payoff desired by Bob. By noting that the second term on the RHS of eq. (\ref{e}) belongs to the stability group of $\ket{\Psi_i}$ we conclude that $U_2\tilde{F}\overline{U}^+_1\overline{V}\tilde{F}^+$ is the relevant counterstartegy to the Alice strategy $V$. \\
As a result, there is no equilibrium in pure strategies unless among $N^2$ classical strategy pairs there exists one leading to an optimal outcome for both Alice and Bob \cite{Landsburg1}. In this sense the pure-strategy equilibria are trivial.

\subsection*{Acknowledgement}
I would like to thank Professor Piotr Kosi\'nski (Department of Computer Science, Faculty of Physics and Applied Informatics, University of L\'od\'z, Poland) for helpful discussion and useful remarks. 
 This research is supported by the NCN Grant no. DEC-2012/05/D/ST2/00754.


\begin{thebibliography}{99}
\bibitem{EisertWL} J. Eisert, M. Wilkens, M. Lewenstein, \emph{Phys. Rev. Lett.} \textbf{83} (1999), 3077
\bibitem{EisertW} J. Eisert, M. Wilkens, \emph{J. Mod. Opt.} \textbf{47} (2000), 2543
\bibitem{Meyer} D. Meyer, \emph{Phys. Rev. Lett.} \textbf{82} (1999), 1052
\bibitem{Marinatto} L. Marinatto, T. Weber, \emph{Phys. Lett}, \textbf{A272} (2000), 291
\bibitem{Benjamin} S. Benjamin, \emph{Phys. Lett.}, \textbf{A277} (2000), 180
\bibitem{MarinattoWeb} L. Marinatto, T. Weber, \emph{Phys. Lett.} \textbf{A 277} (2000), 183
\bibitem{BenjaminHay} S. Benjamin, P. Hayden, \emph{Phys. Rev. Lett.} \textbf{87(6)} (2001), 069801
\bibitem{BenjaminHay1} S. Benjamin, P. Hayden, \emph{Phys. Rev. Lett.} \textbf{87} (2001), 069801
\bibitem{Iqbal} A. Iqbal, A. Toor, \emph{Phys. Lett} \textbf{A280} (2001), 249
\bibitem{DuLi} J. Du, H. Li,X. Xu, X. Zhou, R. Han, \emph{Phys. Lett} \textbf{A289} (2001), 9
\bibitem{EisertWL1} J. Eisert, M. Wilkens, M. Lewenstein, \emph{Phys. Rev.Lett.} \textbf{87} (2001), 069802
\bibitem{FlitneyA} A. Flitney, D. Abbott, \emph{Fluct. Noise Lett.} \textbf{2} (2000), R175
\bibitem{Iqbal1} A. Iqbal, A. Toor, \emph{Phys. Rev.} \textbf{A65} (2002), 052328
\bibitem{DuLi1} J. Du, H. Li, X. Xu, M. Shi, J. Wu, X. Zhou, R. Han, \emph{Phys. Rev. Lett.} \textbf{88} (2002), 137902
\bibitem{Enk} S.J. van Enk, R. Pike, \emph{Phys. Rev} \textbf{A66} (2002), 024306
\bibitem{DuLi2} J. Du, X. Xu, H. Li, X. Zhou, R. Han, \emph{Fluct. Noise Lett.} \textbf{2} (2002), R189
\bibitem{FlitneyA1} A. Flitney, D. Abbott, \emph{Proc. R. Soc. Lond.} \textbf{A459} (2003), 2463
\bibitem{PiotrowskiS} E. Piotrowski, J. Sladkowski, \emph{Int. Journ. Theor. Phys.} \textbf{42} (2003), 1089
\bibitem{DuLi3} J. Du, H. Li, X. Xu, X. Zhou, R. Han, \emph{Journ. Phys.} \textbf{A36} (2003), 6551
\bibitem{Zhou} L. Zhou, L. Kuang, \emph{Phys. Lett} \textbf{A315} (2003), 426
\bibitem{Chen} L. Chen, H. Ang, D. Kiang, L. Kwek, C. Lo, \emph{Phys. Lett.} \textbf{A316} (2003), 317
\bibitem{Lee} C. Lee, N. Johnson, \emph{Phys. Rev.} \textbf{A67} (2003), 022311
\bibitem{Shimamura} J. Shimamura, S. Ozdemir, F. Morikoshi, N. Imoto, \emph{Int. Journ. Quant. Inf.} \textbf{2} (2004), 79
\bibitem{Landsburg} S. Landsburg, \emph{Notices of the Am. Math. Soc.} \textbf{51} (2004), 394
\bibitem{Rosero} A. Rosero, "Classification of Quantum Symmetric Non-zero Sum $2\times 2$ Games in the Eisert Scheme", quant-phys/0402117
\bibitem{NawazT1} A. Nawaz, A. Toor, \emph{Journ. Phys.} \textbf{A37} (2004), 11457
\bibitem{NawazT2} A. Nawaz, A. Toor, \emph{Journ. Phys.} \textbf{A37} (2004), 4437
\bibitem{Iqbal2} A. Iqbal, "Studies in the theory of quantum games", quant-phys/0503176
\bibitem{FlitneyA2} A. Flitney, D. Abbott, \emph{Journ. Phys.} \textbf{A38} (2005), 449
\bibitem{Ichikawa} T. Ichikawa, I. Tsutsui, \emph{Ann. Phys.} \textbf{322} (2007), 531
\bibitem{Cheon} T. Cheon, I. Tsutsui, \emph{Phys. Lett.} \textbf{A348} (2006), 147
\bibitem{Patel} N. Patel, \emph{Nature} \textbf{445} (2007), 144
\bibitem{Ichikawa1} T. Ichikawa, I. Tsutsui, \emph{Journ. Phys. A: Math. and Theor.} \textbf{41} (2008), 135303
\bibitem{FlitneyA3} A. Flitney, L. Hollenberg, \emph{Phys. Lett.} \textbf{A363} (2007), 381
\bibitem{Nawaz3} A. Nawaz, "The generalized quantization schemes for games and its application to quantum information", arXiv:1012.1933
\bibitem{Landsburg1} S. Landsburg, \emph{Proc. Am. Math. Soc.} \textbf{139} (2011), 4413
\bibitem{Landsburg2} S. Landsburg, \emph{Wiley Encyclopedia of operations Research and Management science} (2011)
\bibitem{Schneider2} D. Schneider, \emph{Journ. Phys.} \textbf{A44} (2011), 095301
\bibitem{Schneider} D. Schneider, \emph{Journ. Phys.} \textbf{A45} (2012), 085303
\bibitem{Avishai} Y. Avishai, "Some Topics in Quantum Games", arXiv:1306.0284
\bibitem{Bolonek} K. Bolonek-Laso\'n, P. Kosi\'nski, \emph{Prog. Theor. Exp. Phys.} (2013), 073A02
\bibitem{Ramzan1} M. Ramzan, \emph{ Quant. Inf. Process.} \textbf{12} (2013), 577 \bibitem{Ramzan2} M. Ramzan, M. K. Khan, \emph{Fluctuation and Noise Letters} \textbf{12} (2013), 1350025 
\bibitem{Nawaz4} A. Nawaz, \emph{Chin. Phys. Lett.} \textbf{30(5)} (2013), 050302
\bibitem{Frackiewicz} P. Frackiewicz, \emph{Acta Phys. Polonica B} \textbf{44} (2013), 29
\bibitem{Nawaz5} A. Nawaz, "Werner-like States and Strategies form of Quantum games", arXiv:1307.5508
\bibitem{Nawaz6} A. Nawaz, \emph{J. Phys. A: Math. Theor. J. Phys.} \textbf{45} (2012), 195304
\bibitem{Nash} J. Nash, \emph{Proc. Nat. Acad. Sci.} \textbf{36} (1950), 48; \emph{Ann. Math.} \textbf{54} (1951), 286
\bibitem{Bolonek1} K. Bolonek-Laso\'n, in preparation
\bibitem{Plenio} M. Plenio, S. Virmani, \emph{Quant. Inf. Comput.} \textbf{7} (2007), 1  


\end{thebibliography}
\end{document}